\documentclass{article}

\usepackage[utf8]{inputenc} 
\usepackage[T1]{fontenc}    
\usepackage{hyperref}       
\usepackage{url}            
\usepackage{booktabs}       
\usepackage{amsfonts}       
\usepackage{nicefrac}       
\usepackage{microtype}      
\usepackage{xcolor} 
\usepackage{booktabs}
\newcommand{\RR}{{\mathbb{R}}}

\usepackage{amssymb,amsmath,amsthm}
\usepackage[pdftex]{graphicx}
\usepackage{lmodern}
\usepackage[ruled, vlined,linesnumbered]{algorithm2e}
\usepackage{setspace}
 \usepackage[tiny]{titlesec}
\usepackage{hyperref}
\font\myfont=cmr12 at 16pt

\title{\myfont{ Clustering Molecular Energy Landscapes by Adaptive Network Embedding}}
\author{ Paula Mercurio and Di Liu }

\begin{document}

\maketitle

\begin{abstract}
In order to efficiently explore the chemical space of all possible small molecules, a common 
approach is to compress the dimension of the system to facilitate downstream machine learning 
tasks. Towards this end, we present a data driven approach for clustering potential energy 
landscapes of molecular structures by applying recently developed Network Embedding techniques, 
to obtain latent variables defined through the embedding function. To scale up the method, we also
incorporate an entropy sensitive adaptive scheme for hierarchical sampling of the energy  
landscape, based on Metadynamics and Transition Path Theory. By taking into account the 
kinetic information implied by a system's energy landscape, we are able to interpret dynamical node-node 
relationships in reduced dimensions. We demonstrate the framework through Lennard-Jones (LJ) clusters 
and a human DNA sequence.
\end{abstract}

\section{Introduction}
The motivation of the project is the fundamental question of chemical spaces: how many organic 
molecules can be formed, and of these, how can we identify molecules with useful properties which can be chemically synthesized. Understanding how such molecules
function in biological systems will have a tremendous impact on development of new drugs and new 
treatment of diseases \cite{b:Dobson04}. For example, the GDB-17 dataset \cite{b:Reymond15} takes 
into account only molecules allowed by valency rules, excluding those unstable or unsynthesizable
due to strained topologies or reactive functional groups, thereby reducing the enumeration to a 
manageable database size of 166.4 billion molecules formed of up to 17 atoms of C, N, O, S, and 
halogens. Fast nearest neighbors searching of large generated datasets like GDBs has led to 
methods for virtual screening and visualization of druglike molecules, with early success in 
neurotransmitter receptor and transporter ligands.

Most applications in biology and chemistry, such as protein folding, involve systems that 
behave according to some potential energy landscape of complex structure with a large 
number of local minima, saddle points  (transition states), entropic plateaus and deep energy
wells. Existing methods for energy landscape analysis focus on identifying local minima via 
geometric optimization, and finding transition states connecting them using steepest descent 
pathways \cite{energylandscapes}. To understand the dynamics over multiple
magnitudes of space and time scales, we can take the viewpoint of the system as a network of 
local energy minima and entropic basins, connected by edges weighted according to the energy 
and entropy barriers that must be crossed for transitions between metastable states.

The proposed research is to apply recent Network Embedding techniques 
\cite{deepwalk,node2vec,reactnet,sparseapprox} to develop a data driven approach for {clustering 
of potential energy landscapes} and identifying latent variables of molecular structures, to 
further facilitate sampling and optimization of the chemical spaces and developing generative 
models for druglike small molecules. The latent variables are given by the output of the 
embedding function. By incorporating energetic information, we will be able to interpret 
node-node relationships in reduced dimensions that are consistent with chemical kinetics and are 
more likely to be aligned with synthesizability.

One multiscale challenge is due to the presence of deep potential wells. Metadynamics 
\cite{metadynamics} uses a non-Markovian random walk to explore an energy landscape, which is
smoothed by additive Gaussian terms after each step, and so is the transition probabilities of
the associated random walk. As this happens, the process is discouraged from revisiting the lowest
energy states repeatedly. The eventual output is to have a flattened energy landscape, as well as 
more efficient random walk samplings. The original potential can be recreated by subtracting the 
additive Gaussian terms. 

To study transition processes in complex systems with rugged energy landscape dominated by 
entropic effects, such that transitions involving a flat region on the potential surface that is 
favorable entropically and the necessity to decrease entropy to exit from this region, 
it is necessary to examine the ensemble of all the transition paths as a probability space. The 
Transition Path Theory (TPT) \cite{b:eve06} studies statistical properties of the reactive 
trajectories such as rates and dominant reaction pathways through probability currents between 
adjacent states. In \cite{b:Du}, the definition of probability current in TPT was generalized 
from edges to individual nodes and networks, for characterizing transition states in the form of 
subnetworks.

In this article, we use Network Embedding techniques in combination with Metadynamics and TPT to 
produce adaptive embeddings that hierarchically convey information about the system's behavior at
different scales. We adjust the edge weights of the network in a way that parallels Metadynamics 
to encourage exploration away from the local energy minima, and adopt TPT to capture micro 
dynamical features of interest. It is shown that these embeddings provide an effective way to 
understand and visualize inter-node relationships.

The rest of this article is structured as follows. In Section 2, we provide some background on 
Network Embedding, Metadynamics and TPT. In Section 3, we discuss more details in the 
implementations and demonstrate our method through Lennard-Jones (LJ) clusters. Section 4 
contains an application to a less homogeneous system: DNA folding in a human telomere.

\section{ Background }

\subsection{ Network Embedding }
Real-world networks, particularly those representing possible molecular structures and other
biological and chemical systems, are often large and complex, making them difficult to 
conceptualize. Network Embedding maps nodes of a given network into a low-dimensional 
continuous vector space, to facilitate downstream machine learning tasks. Making use of the 
sparsity of networks, recently developed Network Embedding methods can scale up linearly 
with regard to the number of edges. Major techniques include factorization of 
functions of the adjacency matrix, random walks samplings of node neighborhoods, 
and deep network learning techniques. The idea is that nodes with 
close proximities in the network should have similar embeddings in the latent space. 
 
The basic setup is an undirected network $\mathcal{G}(S,E)$ with node set $S$ and edge set $E$, 
while generalizations to directed graphs is straightforward. Let $|S|=n$ and $A^{n\times n}$ be 
the weighted adjacency matrix of the network with weight $a_{ij}\ge0$ between nodes $v_i$ and 
$v_j$. The output will be a map 
\begin{equation}\label{eq:encode}
z_i=f(v_i): \ S\rightarrow\mathbb{R}^d, \quad\text{with}\quad d\ll n.
\end{equation}
For methods discussed in this paper, the 
encoding function \eqref{eq:encode}, referred as direct encoding, is simply a lookup matrix: 
$z_i=f(v_i) = Ze_i$, such that $Z\in\mathbb{R}^{d\times n}$ contains the embedding vectors for 
all nodes $v_i\in S$, and $e_i$ is an indicator vector, therefore $f(v_i)$ simply gives the $i$th 
column of $Z$. The set of trainable parameters for direct encoding approaches is the embedding 
matrix $Z$, which is to be optimized directly. Vector $a_i=\{a_{ik}\}_{k=1}^{n}$ denotes the first
order proximity between node $v_i$ and other nodes. The second order proximity between $v_i$ and 
$v_j$ can be determined by the similarity between $a_i$ and $a_j$, which compares the pair's 
neighborhood structures. 

In DeepWalk and Node2vec \cite{deepwalk, node2vec}, short random walk simulations are used to 
determine proximity for each pair of nodes. More specifically, random walk runs through nodes of 
the network, with transition rates determined by the edges weights. Two nodes are close to each 
other if there is a high probability that a random walk simulation containing one node will also 
contain the other. Similarities between between embedding nodes is given by the following 
conditional probabilities based on the SkipGram model:
\begin{equation}\label{skip}
 P(j|i) = \frac{\exp(z_i \cdot z_j)  }{\sum_l \exp(z_i\cdot z_l)},
\end{equation}
where $P(j|i)$ denotes the conditional probability that a random walk starting at node $v_i$ will 
include node $v_j$, and $z_i$ is the embedding of $v_i$. The learning is achieved by minimizing 
the following cross entropy loss using Stochastic Gradient Descent (SGD) method:
\begin{equation}\label{eq:loss}
\mathcal{L}(\mathcal{G}) = \sum_{v_i} \sum_{j \in R(v_i)} -\log\left( P(j|i) \right),
\end{equation}
where $R(i)$ represents a $k$-step random walk trial starting from node $i$.
The efficiency of evaluating \eqref{skip} can be greatly improved by using negative sampling 
\cite{negsampling} that randomly selects edges favoring less frequent ones.

In this paper, we will adopt a Network Embedding scheme introduced in \cite{sparseapprox}, where
the embeddings are produced via a sparse approximation of random walks on networks. For a given 
undirected network $\mathcal{G}(A)$ with adjacency matrix $A=(a_{ij})$, let $D$ be the diagonal 
matrix such that $D_{ii}=\sum_j a_{ij}$. The volume of the graph will be given by 
$v = \sum_iD_{ii}$. The Laplacian $L=I-D^{-1}A$ has eigen-decomposition $L=\Phi\Lambda\Phi^T$, 
where $\Lambda$ represents the diagonal matrix of ordered eigenvalues so that 
$0=\lambda_1 \leq\lambda_2\leq...\lambda_{n}$, and the eigenvectors are given by columns of 
$\Phi$ denoted by $\phi_1, \phi_2, ....\phi_{n}$. Assuming the network is connected, the discrete 
Green function satisfies
\begin{equation}
G(i,j)=\sum_{k=2}^{n}\frac{1}{\lambda_k}\phi_k(i)\phi_k(j).
\end{equation}
We can further define the commute time $ct(i,j)$ to be the mean time for the Markov process 
prescribed by transition probability matrix $T=D^{-1}A=I-L$ on the network to travel from node 
$v_i$ to node $v_j$, and back to $v_i$. It is shown in \cite{qiu} that the coordinate matrix for 
embeddings that preserve commute times has the following form
\begin{equation}\label{eq:theta}
\Theta = \sqrt{v}\Lambda^{-1/2}\Phi^T.
\end{equation}

To produce an efficient approximation of $\Theta$, we can assume that $T$ is local in the sense 
that, at least asymptotically, its columns have small support, and high powers of $T$ will be of 
low rank, which can be justified for potential driven systems by disparate transition rates 
between different neighboring metastable states. Taking higher powers of $T$ is equivalent to 
running the Markov chain forward in time, which allows for representations of the random walk,
i.e. reaction pathways, at different time scales. Making use of this sparsity, we can produce 
compressed approximations to the dyadic powers of $T$ with its principal components, by using 
fast algorithms such as Lanczos Bidiagonalization for singular value decomposition (SVD) with 
the complexity depending linearly on the number of nonzero elements. 

The following scheme is a modification of the Diffusion Wavelet algorithm \cite{diffwavelets}.
Starting with $T_0=T$, at each iteration, taking $U_k$ and $\Sigma_k$ to be the top left singular 
vectors and singular values of $T_{k} \approx T^{2^{k}}$, we let $T_k := U_k^T T_{k-1} U_k$. From 
this, we can have a low-rank approximation to the Green function of the random walk, using the 
Schultz method:
\begin{equation}
{G} = \sum_{k=1}^\infty T^k  =  \prod_{k=0}^\infty(I + T^{2^k}) .
\end{equation}
The embedding matrix $\Theta$ thus satisfies $\Theta^T \Theta = v G$, where $v$ is the volume of 
the network defined as above. Denoting the leading singular values and left singular vectors of 
the matrix $v{G}$ by $\Sigma_G$ and $U_G$, we take $\Theta := \Sigma_G^{1/2}U_G^T$. 

We can further use $\Theta$ as a starting point, introduce parameters by multiplying its $j$th
column by a weight $c_j$, and optimize $\{c_j\}$'s to minimize cross entropy loss 
\eqref{eq:loss} via SGD. Moreover, for robustness of the algorithm, with certain probability, we 
can reintroduce singular vectors that were removed in previous truncations, as a residual 
correction technique.

\subsection{ Metadynamics}
Metadynamics was introduced in \cite{metadynamics} as a technique to aid in the exploration of 
energy landscapes. The scheme is to create a non-Markovian, and approximately self-avoiding,
random walk by adjusting the gradient of the energy landscape after each step with the addition 
of derivative of a Gaussian term. Over time, these Gaussians eventually fill up the valleys in 
the energy potential, which allows the random walk to explore other areas of the landscape and 
leads to a more complete picture of the system dynamics. Specifically, after each step, the 
parameter $\phi_i$, which represents the derivative of the energy with respect to the $i$th 
parameter $-\frac{\partial E}{\partial x_i}$, is adjusted according to:
\begin{equation}
\phi_i^{t+1} = \phi_i^t - \frac{\partial}{\partial x_i} W 
\prod_i \exp \left(-\frac{|x_i-x_i^t|^2}{2\delta^2}\right),
\end{equation}
where $W$, $\delta$ and $x_i^t$ are the height, width and center of the Gaussian respectively, 
to be chosen based on prior knowledge of the energy landscape.

\subsection{ Transition Path Theory}
The Transition Path Theory (TPT) \cite{b:eve06} studies statistical
properties of the reactive trajectories such as rates and dominant pathways through
probability currents between adjacent states. In the simplest setting, given reactant state 
$A$ and product state $B$, any equilibrium path $X(t) $ oscillates infinitely many times 
between $A$ and $B$, with each oscillation from $A$ to $B$ being a reaction event. The 
reactive trajectories are successive pieces of $X(t) $ during which it has left $A$ and on 
its way to $B$ next, without coming back to $A$. 

The discrete forward committor $ q^+_i $ is defined as the probability that the process 
starting in node $i$ will first reach $B$ rather than $A$, and the discrete backward 
committor $ q^-_i $ is defined as the probability that the process arriving in node $v_i$ last 
came from $A$ rather than $B$. For Markov processes with infinitesimal generator 
$T=(t_{ij})$, the forward committor satisfies discrete Dirichlet equations:
\begin{equation}\label{eq:commitor}
\sum\limits_{v_j\in S}t_{ij} q_j^+=0, \quad \text{for}\ v_i\in S\backslash(A\cup B),  
\end{equation}
with the boundary condition $\left(q_i^+=0, \text{if}\ v_i\in A; 
\ q_i^+=1,\text{ if }\ v_i\in B\right)$.
The backward committor satisfies a similar equation. For time reversible processes, 
$q_i^+=1-q_i^-$. 

The probability current of reactive trajectories is the average rate at which they flow from 
one state to another when the process is at statistical equilibrium with distribution $\pi$, 
and can be obtained by
\begin{equation}\label{eq:current}
f^{AB}_{ij}=\pi_iq^-_it_{ij}q^+_j,  \textrm{\quad if $i\neq j$}.
\end{equation}
To deal with the fact that transitions between any two states can go forward and backward, 
the effective current can be introduced as 
$f^{+}_{ij}=\max\big(f^{AB}_{ij}-f^{AB}_{ji},0\big)$.

\section{Network Embedding with Metadynamics}
In this section, we want to introduce the Network Embedding techniques for energy landscapes with 
Metadynamics adjustments through a few of examples. 

\subsection{ 8 atom Lennard-Jones cluster}
LJ clusters are often used as a model for atomic or molecular dynamics within a fluid,
in which the potential energy $E$ of a given configuration of atoms depends only on distances 
between atoms:
\begin{equation}
E(r) = 4\epsilon 
\sum\left[ \left( \frac{\sigma}{r_{ij}}\right)^{12}-\left(\frac{\sigma}{r_{ij}}\right)^6\right],
\end{equation}
where $r_{ij}$ denotes the distance between atoms $i$ and $j$, and the parameters $\sigma$ and 
$\epsilon$ represent pair equilibrium separation and well depth. In the experiments below, we 
adopt reduced units (e.g. $\sigma=\epsilon=1$).

\begin{figure}[ht] 
    \includegraphics[width=.45\linewidth]{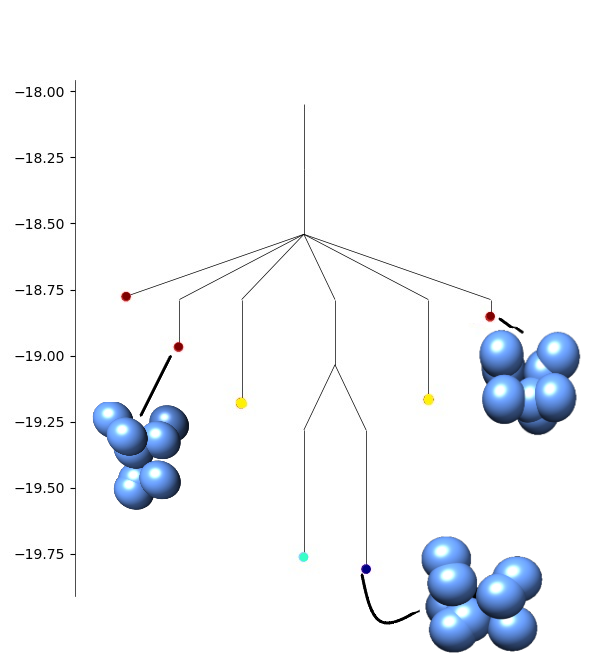}  
    \hfill
    \includegraphics[width=.45\linewidth]{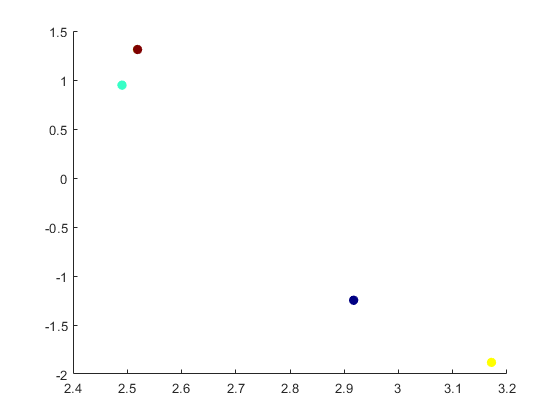} 
   \caption{Disconnectivity tree and Metadynamics based embeddings for the Lennard-Jones cluster 
   with 8 atoms. Left: Disconnectivity tree of all local minima. Right: Embeddings for the local 
   minima after applying the Metadynamics adjustment. 
   Color scheme represents the potential energy, e.g., dark blue 
   denotes the lowest, and red as the highest.
   Closely related minima have very similar or identical embeddings, e.g., both yellow 
   minima are embedded at the yellow point on the right.  
   \label{fig:lj8V}} 
\end{figure}

To apply the Network Embedding techniques to the LJ clusters, we first generate a 
database of local energy minima using the Pele software available at \cite{walessite}. The 
database was produced using a basin-hopping run of 500 steps, and consists of 8797 local minima 
connected by 8099 transition states. The database also contains thermodynamic information, 
including the potential energy at each local minimum. The embeddings here are based on a network 
constructed with nodes given by the local minima, and edges located between each pair of nodes 
where a transition state has been identified. The adjacency matrix has entries given by the
energy barriers between metastable states. 

Initially, every node is embedded into the vector space. Since most points near the global 
minimum will be close to each other in terms of commute time, this typically results these nodes 
being embedded around the global minimum. Then, removing nodes that are further away from
the global minimum, and re-embedding only the nodes in residual cluster reveals new, more 
detailed information about the remaining nodes. We do this by creating a new, smaller adjacency 
matrix including only re-embedded points, and adjusting the edge weights with Gaussian terms 
according to Metadynamics 
\begin{equation}\label{adjmetadynamics}
a_{ij}^{t+1}= a_{ij}^{t} - W \prod_i \exp(-\frac{||\theta_i - \theta_j||^2}{2\delta^2}),
\end{equation}
where $\theta_i$ is the coordinate of the $i$th node (energy minimum). Adaptively choosing the
center node and removing distant nodes, the process will produce a series of hierarchical 
embeddings that provide information about the full energy landscape. Each level provides a 
representation of the energy landscape at a different scale. 

Figures \ref{fig:lj8V} gives the disconnectivity tree of the original potential and embeddings
after applying the Metadynamics adjustment by equation \ref{adjmetadynamics} for the 8-atom
LJ cluster. The colors on each minimum on the disconnectivity tree are matched to those
nodes' embeddings in Figure \ref{fig:lj8V}, with some of the nodes colored red,
embedded to the same place. In the embeddings, nodes with closer dynamic relationships are
clustered together, as a result of the shorter commute time distance between them. 

On a larger scale, we can also see how the higher energy nodes relate to the nodes with the two 
lowest energies. Specifically, the global minimum (in dark blue) is closer to the nodes in the
middle of the tree (in orange), while the second lowest local minimum is much closer to the three
highest energy nodes. This allows us to draw conclusions about a lowest commute time path through 
these states: one of the orange colored states might transition directly to the global minimum, 
while one of the higher energy states would be more likely to transition to the second lowest
energy state first, and then either remain there or transition on to the global minimum. 

\begin{figure}[ht]
\centering
\includegraphics[width=0.5\linewidth]{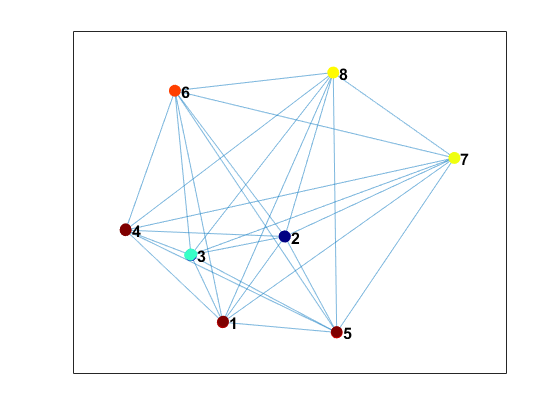}
\caption{The 8-atom LJ network of local minima. Edge lengths are proportional to 
commute times. Node colors are chosen to match those in Figures \ref{fig:lj8V}. 
\label{fig:lj8graph}}
\end{figure}

\begin{table}[h]
  \begin{center}
    \caption{Commute times between nodes, 8-atom LJ cluster.     \label{tab:lj8}}

    \begin{tabular}{lcccccccr}\toprule 
    \textbf{Node 1} & \multicolumn{8}{c}{\textbf{Node 2}}
\\\cmidrule(lr){2-9} \\
           & 1   & 2  & 3   & 4  & 5 & 6 & 7 & 8 \\
      \midrule
1 & 0 & 25.4 & 17.7 & 23.5 & 29.0 & 65.2 & 86.1 & 88.4 \\
2 &  & 0 & 7.7	& 13.5 & 19.0 & 55.2 & 60.7 & 63.0 \\
3 &  & &	0 &	5.8 &	11.3 & 47.5 & 68.4 & 70.7 \\
4 &  &	 &	 &	0 &	17.1 &	53.3 &	74.3 & 76.5 \\
5 &  &	 &	 &	 &	0 &	58.8 &	79.7 & 82.0 \\
6 &  &	 &	 &	 &	 &	0 &	115.9 & 118.2 \\
7 &  &	& & & & &	0  & 123.7\\
     \bottomrule
    \end{tabular}
  \end{center}
\end{table}

For comparison, we also investigated the inter-node relationships by directly computing the 
commute times between each pair of nodes. The commute time between nodes $v_i$ and $v_j$, or mean 
time for the Markov process on the graph prescribed by the Laplacian $L$ to travel from $v_i$ to 
$v_j$ and back to $v_i$, is given by 
\begin{equation}    
ct(i,j) = v\sum_{k=2}^N \lambda_k (\phi_k(i)-\phi_k(j) )^2,
\end{equation}
where $\lambda_k$ are the eigenvalues of $L$, and the $\phi_k$ are the corresponding eigenvectors
as defined in the introduction. The off-diagonal entries for the Laplacian for a LJ cluster with 
$\kappa$ degrees of freedom are given by 
\begin{equation}
L(i,j)= \sum_k \frac{O(i)}{O_k(k)}\frac{v(i)}{v_k(k)}^{\kappa-1}v(i)\exp(-\beta(E_k(k)-E(i))),
\end{equation}
where $O$ and $O_k$ represent point group orders of the local minima and transition states, 
respectively, similarly $v$ and $v_k$ represent mean vibrational frequencies and $E$ and 
$E_k$ represent potential energy levels at each configuration. It can be seen that the commute 
time diagram is consistent with the Network Embedding output.

\subsection{Multi-level embeddings of LJ 38 cluster}

\begin{figure}[t] 
\centering
 \includegraphics[width=.5\linewidth]{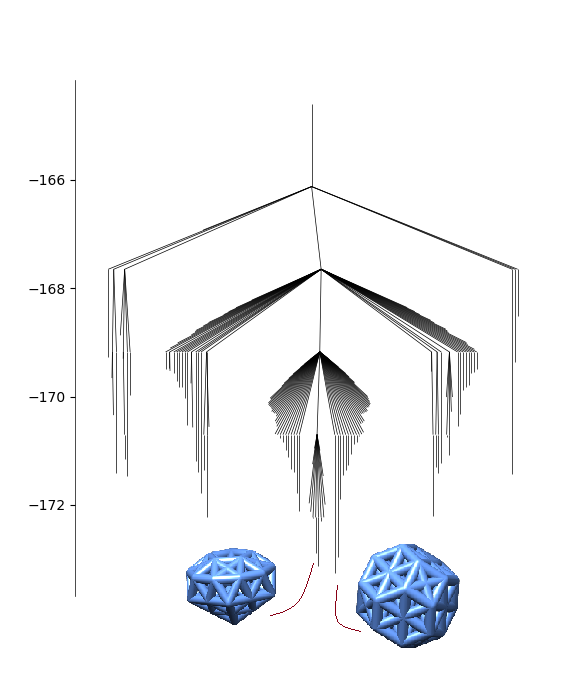} 
    \caption{ Disconnectivity tree for the 38-atom LJ cluster. 
    The structures of the two lowest-energy configurations are also pictured.
    \label{fig:lj38tree}}
\end{figure}

\begin{figure}[t] 
 \includegraphics[width=.5\linewidth]{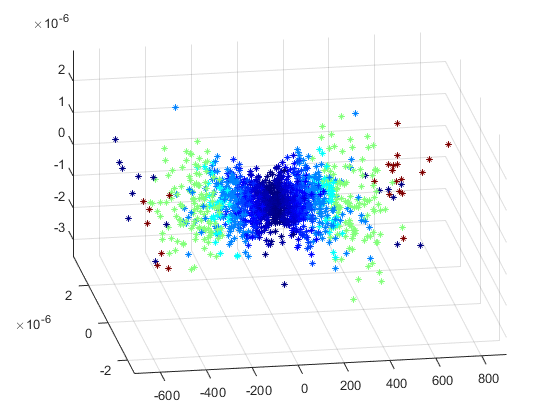}  
    \includegraphics[width=.5\linewidth]{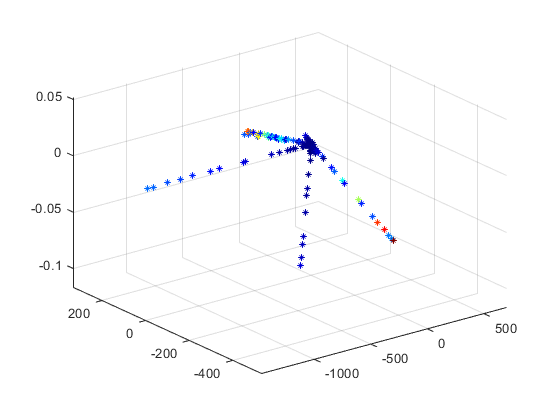} 
    \caption{Hierarchical embeddings for the LJ cluster with 38 atoms. 
    Pictured are the embeddings before (left) and after (right) applying Metadynamics.   
    Color scheme denotes commute time from the global minimum, 
    with dark blue being shortest distances, and red as furthest distances. 
    \label{fig:lj38}}
\end{figure}

Figure \ref{fig:lj38tree} shows disconnectivity tree of the 38-atom LJ cluster. 
For simplicity, we construct the adjacency matrix for this network with entries given by the 
energy barriers between states. Figure \ref{fig:lj38} gives the hierarchical embedding of the 
LJ-38 cluster. The left image shows the initial embeddings, colored by their commute time 
distances from the global minimum. We re-embedded the points of interest with the Metadynamics
adjustment to the potential with two iterations and obtain the right image. 

The hierarchical structure of the embedding process is organized as the following. In the first 
level embeddings (Fig. \ref{fig:lj38}, left), the nodes are embedded consistently with their 
commute time distances from the global minimum, which correspond loosely to potential energy 
level. In particular, the nodes with energy $E>-170$, which are the highest energy clusters in 
the tree, are embedded further from the global minimum. In the next embedding, these nodes are
removed due to their further distances from the global minimum, and more central clusters will
be re-embedded. As we consider the second level and later embeddings, this pattern repeats: 
each re-embedding reveals a new ``layer" of nodes that are embedded closer to the global 
minimum, which correspond to nodes on the disconnectivity tree that are of lower energy than  
nodes removed in the previous level.

In particular, at the 3rd level (Fig. \ref{fig:lj38}, right), the embeddings have 4 small 
``spokes" originating from a central cluster containing the global minimum. However, these
embeddings provide additional context: nodes that are embedded within a particular ``spoke" are 
more closely related, which means the system is more likely to transition between these states. 
Since the nodes do not all come from the same group on the disconnectivity tree, these embeddings 
can also reveal interactions between nodes that aren't indicated on the disconnectivity tree. 

Additionally, since the spokes are connected to the cluster containing the global minimum, we can 
conclude that each spoke represents a potential transition pathway from the outer edge of the 
cluster to the center. In other words, if the system is at a state represented by the outer point 
of one of the spokes, its most likely path toward the global minimum will be to travel through 
the states represented by other nodes in the same spoke.

We can apply similar reasoning to higher level embeddings. Each level reveals a more detailed 
picture of the dynamics of a different part of the system's energy landscape. The first levels 
give a coarse-grained picture, only identifying broad groups of high energy and low energy nodes,
while later levels give a more fine-grained visualization of the nodes most closely related to 
the global minimum.

\begin{figure}[ht] 
  \centering
    \includegraphics[width=.5\linewidth]{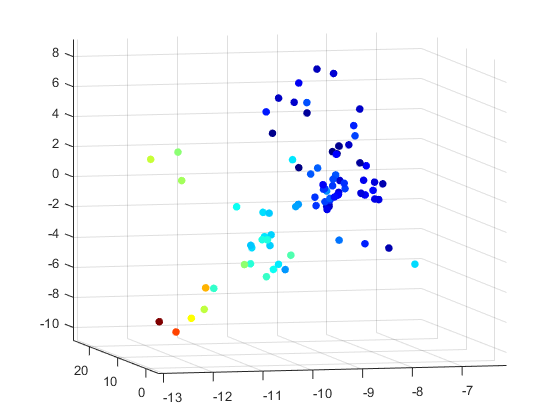} 
    \caption{Embeddings of the local minima of the 38-atom LJ cluster with potential 
    energies less than -170.9. The figure shows output of 2 level embeddings with Metadynamics 
    adjustment. Color scheme denotes commute time from the global minimum.
    \label{fig:ljtrunc}}
\end{figure}

Often, we want a more detailed, finer grained visualization of the energy landscape than the 
disconnectivity tree in Figure \ref{fig:lj38tree} can provide. It is informative, therefore, to 
re-embed parts of the network of greatest interest to gain further insight. Here we focus on 
the lowest energy parts of the LJ energy landscape. We repeated the above process using the 
subnetwork consisting only of the nodes with potential energy $<-170.9$, that is, the 163 lowest
energy nodes. Figure \ref{fig:ljtrunc} shows the results of this experiment. 

Now we want to provide a closer inspection on the information flow along the hierarchical 
sampling with Metadynamics. In the first level without the Metadynamics adjustment, 
nodes in these embeddings (both Figure \ref{fig:lj38} and Figure \ref{fig:ljtrunc}) are clustered 
according to their similarity in terms of commute time. In other words, nodes that can be quickly
and frequently reached from either the global minimum or second lowest energy node will be 
grouped near them. As a result, most of the nodes we are most interested in end up in the same 
cluster, and the embeddings obtained from the first application of the embedding method only give
useful clusters for nodes that are more distant in terms of commute time from the part of the 
graph of interest. As we progress through additional levels, we pull apart the cluster 
containing the global minimum, positioned near the origin in the first level's embeddings, 
until the final level's embeddings give details of the dynamics of the process within this 
cluster. 

After the second level of Embedding with Metadynamics adjustment, if two nodes share a 
cluster or are close in the embedding space, it indicates that the system can easily transition
between those nodes, with a relatively low energy barrier. As a result, the groupings seen in 
these embeddings correspond to the groupings in the disconnectivity tree for this system 
\cite{energylandscapes, walessite}. For instance, one of the clusters in the second level 
embeddings correspond to the global minimum and its nearest neighbors, pictured directly right of 
center in the disconnectivity tree in Figure \ref{fig:lj38tree}. Clusters can be mapped to the 
disconnectivity tree by comparing the potential energies of the nodes within the cluster to the 
tree.

The difference is after the third level, some of the clusters instead represent combinations 
of multiple disconnectivity tree groups; this is a result of re-embedding the nodes to spread out 
those that were previously near the origin. Such nodes ended up being embedded in or near 
the clusters they are most closely related to, even though they are not actually members of the 
respective tree groupings. For instance, the global minimum is embedded directly next to a node
from a neighboring tree group.

In other words, the first level's embeddings tell us about higher energy nodes and those that 
are more distant from the global minima, and further levels reveal information about parts of 
the network that are closer to the global minimum. Additionally, these embeddings are useful for 
identifying transition paths. The structure of the third level embeddings, in particular, reveals
four transition paths: if the system is initialized from a node near the outside of one of these 
``spokes", its lowest energy path to the global minima will involve following the spoke into the 
center cluster. 

\begin{figure}[ht] 
  \begin{minipage}[b]{1\linewidth}
    \includegraphics[width=.5\linewidth]{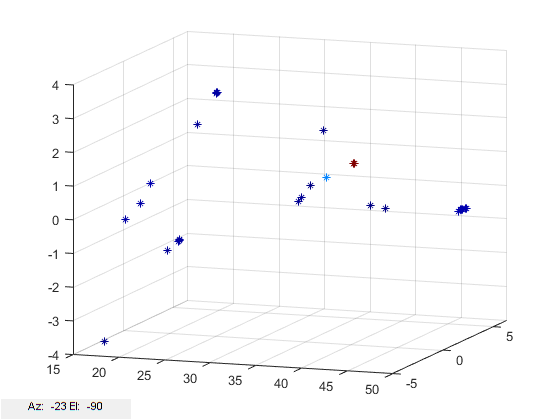}  
    \hfill
    \includegraphics[width=.5\linewidth]{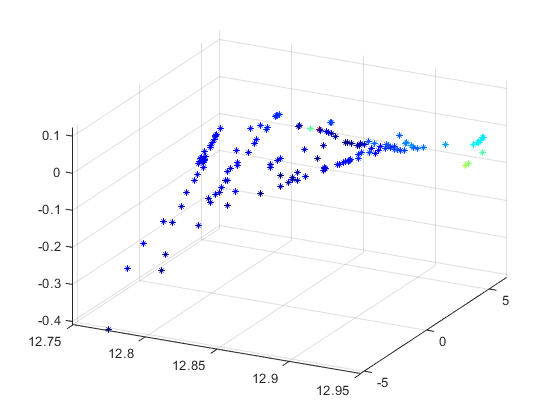} 
       \caption{Metadynamics-based embeddings for the 38-atom cluster at the temperatures 
       $T=0.08$ (left) and $T=1$ (right). Color scheme denotes commute time from the global 
       minimum. \label{fig:lj38entropy}} 
  \end{minipage}
\end{figure}

We can also use these embeddings to observe the results of entropic changes to the system. In 
Figure \ref{fig:lj38entropy}, the embeddings for this cluster under two additional temperature 
conditions are shown. The results of the temperature change are reasonable according to what was 
observed previously with the 8-atom cluster. Namely, at lower temperatures 
(Fig.\ref{fig:lj38entropy}, left) closely related nodes are more likely to be embedded much nearer
each other, creating the impression that there are fewer embeddings, while at higher temperatures 
(Fig.\ref{fig:lj38entropy}, right), there is greater variation in the node embeddings. 

\subsection{Human Telomere Folding}
Now we want further develop Network Embedding technique for organic molecular structures and 
apply it to a more complex problem: DNA folding in a human telomere. More specifically, we 
consider a sequence of 22 nucleotide bases $A(G_3TTA)_3G_3$ which repeats within human telomeres. 
This sequence is known to form a G-quadruplex, a type of secondary structure formed by groups of 
four guanine bases called G-tetrads. Its structure and potential energy landscape were previously 
investigated in \cite{DNAGquadruplex}, where the potential energy landscape was calculated using 
the HiRE-RNA model for coarse-grained DNA \cite{hirerna} with 6 or 7 atoms considered for each of 
the 22 nucleotides in the telomere. We use this database as a starting point. In particular, we 
construct a network with nodes given by the 4000 lowest-energy local minima, and edges between 
nodes determined by the transition states connecting them. 

\begin{figure}[t] 
  \begin{minipage}[b]{1\linewidth}
   \centering
    \includegraphics[width=.5\linewidth]{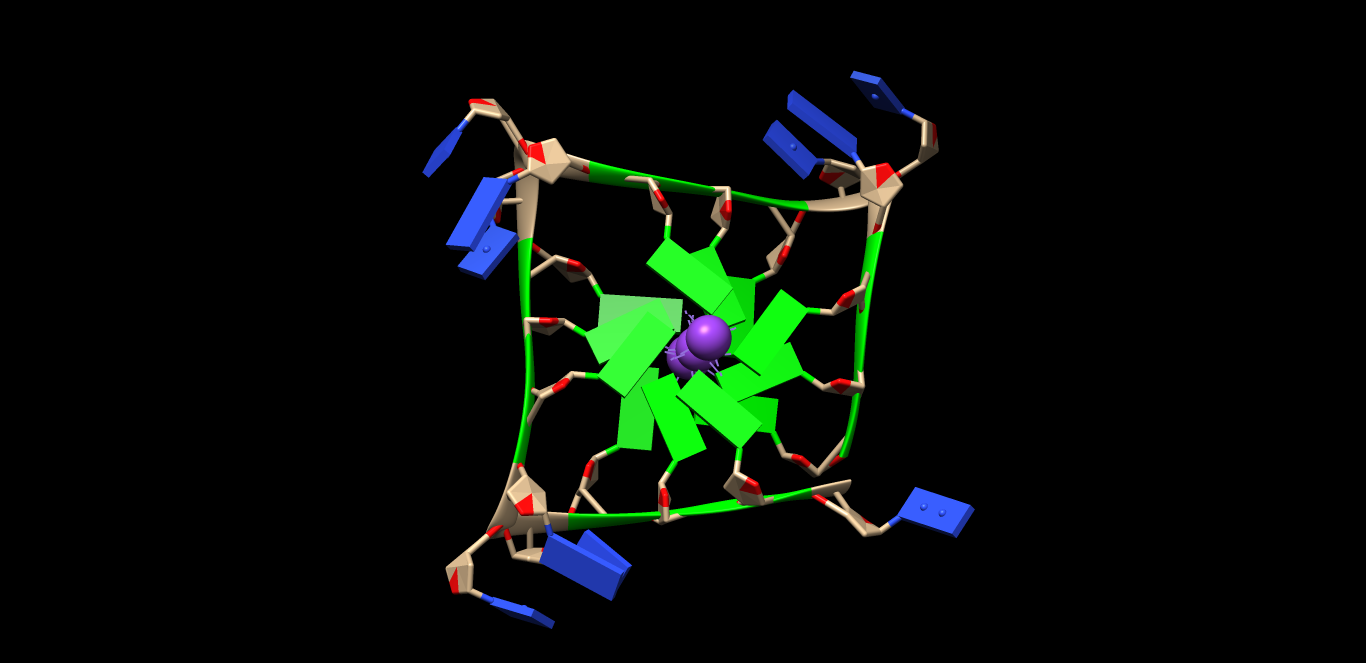}  
    \caption{The four-strand G-quadruplex structure (PDB structure 1KF1), with guanine 
    nucleotides colored green. Image produced with Chimera \cite{chimera}.\label{fig:quadruplex}} 
  \end{minipage} 
\end{figure}

\begin{figure}[ht] 
  \begin{minipage}[b]{1\linewidth}
    \includegraphics[width=.5\linewidth]{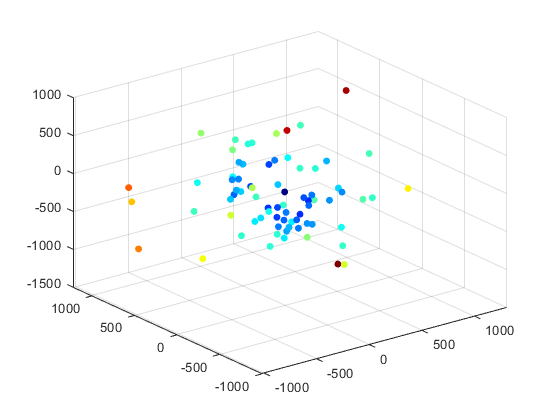} 
    \includegraphics[width=.5\linewidth]{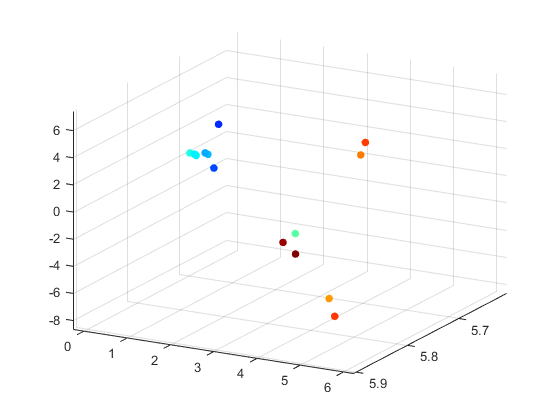} 
    \caption{ Embeddings of the local minima network for the human telomere sequence, based on 
    the adjacency matrix and the Metadynamics adjustment.  
    Color scheme denotes commute time from the global minimum.
    \label{fig:telomere}}
  \end{minipage} 
\end{figure}

For first experiments, we used a random walk based on the energy barriers between states. Figure 
\ref{fig:telomere} shows the results of the initial embeddings and fourth 
levels with metadynamic adjustments, based on an energy landscape adjusted by a Gaussian 
term with width 0.75 and height 1 after each successive embedding. As in the LJ cluster
experiment, the first embedding includes all nodes in the network, while each successive embedding
shows a re-embedding of the subnetwork of nodes most closely related to the global minimum (that
is, all nodes whose previous embeddings lie within a small tolerance of the global minimum).

As with the LJ clusters, each level of embedding represents ``zooming in" on the part 
of the network around the global minimum. The first level gives us a global view -- the global 
minimum and its nearest neighbors (with respect to commute times) are clustered at the origin, 
represented by a dark blue dot. The red and orange dots furthest away from the origin represent 
local minima which are more distantly related, requiring multiple steps or higher energies to 
transition to the global minimum. Potential transition paths can be identified by starting at one
of these points, and moving toward the origin along nearby points. At the second and each 
of the following levels, the nodes embedded closest to the local minimum are re-embedded to give
us a more detailed inspection into the relationships of those nodes. The likely transition 
paths here can be constructed similarly.

\section{Multiscale embedding with TPT}

Now we want to make use of the multiscale nature of the molecular dynamics to speed up and scale
up the computation. When a subset of the variables evolves more quickly than the others, 
dimension reduction can be achieved by the quasi-equilibrium on the fast variables such that 
averaged macroscopic effect can replace microscopic details. For molecular configurations where 
the energy landscape is ``flatter" and state transitions occur faster, this principle of averaging 
applies. From the space perspective, if the transitions paths between nodes representing 
different molecular configurations are of relatively low energy barriers, they should be 
closely related in terms of mean commute time. We can expect these low barrier transitions
to describe subtler changes likely involving position changes for only a small number of atoms.
In other words, there is a time-space scale separation, i.e., on a global scale, transitions 
between two states tend to require larger, higher dimensional changes (associated with greater 
energy expenditures), while locally, transitions within some subnetwork around a point of 
interest require far fewer degrees of freedom. 
 
For energy landscapes with a large number of local minima, the corresponding networks contain 
large numbers of nodes. In fact, for the LJ clusters, the number of minima increases 
exponentially with the number of atoms. In these situations, as we have seen, it is benefiting to 
have a method for Network Embedding that focuses on locally embedding regions of the network that 
are of particular interest, for example the subnetwork consisting of the global minimum and its 
nearest neighbors. Since the commute times between states in this subnetwork are short, the time 
required for the system to move between these configurations is fast and the nodes tend to 
embedded near each other. The states in these subnetworks often represent molecular 
configurations that are similar, differing only by a simple conformational change; therefore 
these subnetworks can have dramatically reduced dimensions compared with the full network.

We want to present an alternate formulation, which replaces the adjacency matrix with the 
probability current matrix from TPT. For large networks, particularly those with irregular 
structures, the committor functions needed to apply TPT to the full network may prove difficult 
or impractical to compute. Focusing our application of TPT onto a smaller, localized subnetwork 
avoids this difficulty, allowing us to take advantage of the additional information TPT offers.
In the following experiments, we first embed all the local minima using the adjacency matrix
constructed with energy barriers, and apply the Network Embedding with Metadynamics. At each 
level, we remove nodes that have longer commute times to the global minimum, until the 
system is reduced to a subnetwork such that we can apply TPT and compute the committor functions  
\eqref{eq:commitor} and probability currents \eqref{eq:current}.
                  
Then, each subnetwork is embedded into $\RR^3$ using the effective probability current in place
of the adjacency matrix. Here we apply the hierarchical embedding procedure to the subnetwork of 
nodes surrounding the global minimum as in previous sections, but the same 
process could be used to examine other domains of the network aside from the global minimum to 
obtain a complete picture of the energy landscape.

The distances between nodes within each subnetwork preserve the commute times. Since the cross 
entropy loss minimization is applied to the overall network at each step, we can expect that the
distances between node embeddings to be consistent with likelihoods of a transition, similarly 
to the previous examples, and therefore we can interpret the embeddings and identify possible 
transition paths in the same way.

\begin{figure}[ht]  
  \begin{minipage}[b]{1\linewidth}
  \centering
    \includegraphics[width=.5\linewidth]{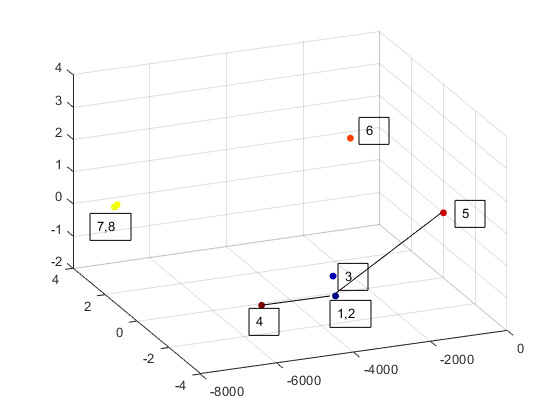} 
    \caption{ Hierarchical embeddings of the local minima network for the 8 atom LJ cluster, 
    based on a TPT-based subnetwork consisting of the nodes clustered around the global minimum 
    (in dark blue), and the metadynamic adjustment. Colors of embeddings denote potential energy 
    same as in previous illustrations. 
    \label{fig:localtptlj8}}
  \end{minipage} 
\end{figure}

We can demonstrate its efficacy on a smaller system--the 8-atom LJ cluster, as 
illustrated in Figure \ref{fig:localtptlj8}. We still see that the two nodes with potential 
energies near -19.2 (colored yellow in Figs \ref{fig:lj8V} and \ref{fig:lj8graph}) are closely 
connected, and the nodes represented in red and orange are closer to the lowest energy nodes than 
to each other, but in this case the short distance between the 2 lowest energy nodes reflects a 
higher transition rate between them. The highest energy nodes, embedded in red, are also embedded 
separately in this case, whereas their adjacency matrix based embeddings were identical.
These embeddings indicate that a transition path (shown in Fig. \ref{fig:localtptlj8}) from the 
highest energy node to the slightly lower energy node labeled 5 might pass through nodes 1 and 2 
on the way. Direct simulations of this system confirm this transition path. Hence, these 
embeddings can be useful for predicting mechanisms by which a molecule changes between two 
configurations.

\begin{figure}[ht] 
    \includegraphics[width=.5\linewidth]{tel_tpt_1_3d.png} 
    \hfill
   \includegraphics[width=.5\linewidth]{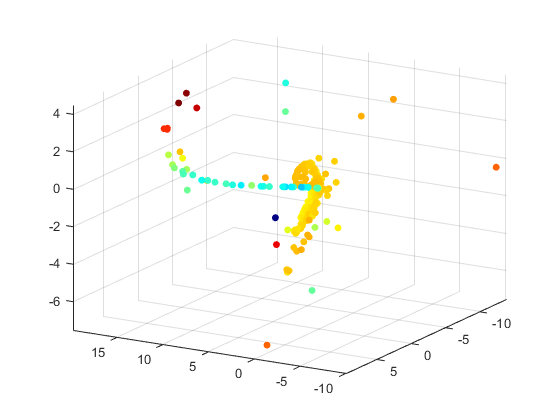}
    \caption{ Hierarchical embeddings of the local minima network for the human telomere 
    sequence. Left: the embeddings of the full network based on the adjacency matrix. 
    Right: embeddings of the subnetwork near the global minimum using Metadynamics and TPT.
    Colors of embeddings denote commute time distance from the global 
    minimum.\label{fig:telomere2} }
\end{figure}

We now return to the human telomere molecule. Figure \ref{fig:telomere2} shows the embeddings 
produced via the TPT process described above. The colors of local minima in Figure 
\ref{fig:telomere2} are determined by the commute times between those nodes and the global 
minimum, which is embedded in dark blue. It is immediately apparent that the local minima are 
grouped according to these commute times, with separate clusters containing most of the red, 
yellow, and light blue nodes. Within each of these clusters, the nodes are relatively close in 
terms of commute times, indicating that these molecular configurations are similar up to some 
simple molecular change.  As with  the 8-atom LJ cluster, these embeddings suggest possible 
transition paths. For example, if the system starts at one of the states furthest from the global 
minimum (colored in red, clustered in the upper left of the right figure) one would expect a 
transition path to the global minimum to travel through the light blue and yellow clusters to 
reach the node in dark blue. 
It is also worth noting that the embeddings given by the TPT approach appear to retain more nodes compared to the adjacency matrix-based embeddings in Figure 8, which results from the greater number of nonzero edge weights in the probability current matrix.

\section{Conclusion and future work}
This article presents a framework for the analysis of energy landscape data. Adaptive network 
embeddings that combine the ideas of Metadynamics and TPT with Node Embedding techniques can be
used to aid in the interpretation and simplification of energy landscape data. In this embedding 
scheme, the network itself -- both its edge weights and the set of nodes under consideration -- 
can be adjusted to more effectively focus on particular areas of the graph. Future research will 
involve applying and developing the method for functions of specific molecules in the context of
certain chemical or biological reacting networks. 

We anticipate that these energy landscape-based network embeddings would be used to advance the 
models currently used to explore the space of small molecules and identify potential new drugs.
For example, the latent variables learned from molecular energy landscapes could be incorporated 
into Variational Autoencoders or other generative models as node attributes 
\cite{gomez-bombarelli}, in much the same way that 3D representations of molecules are 
already used. The inclusion of these additional latent variables would lead to a multi-modal 
generative method that takes into account kinetic information of the molecular systems for 
generating more realistic, chemically viable molecules.

%

\section{Declarations}

\subsection{Acknowledgments}
The work is partially supported by NSF-DMS 1720002.

\subsection{Authors’ contributions}


Made substantial contributions to conception and design of the study and performed data analysis and interpretation: Mercurio P

Made substantial contributions to conception and design of the study and performed data analysis and interpretation, as well as provided administrative and material support: Liu D

\subsection{Availability of data and materials}

Molecular graphics and analyses performed with UCSF Chimera \cite{chimera}, developed by the Resource for Biocomputing, Visualization, and Informatics at the University of California, San Francisco, with support from NIH P41-GM103311. 

Databases of local minima and transition states were computed using the Pele software \cite{pele} and PATHSAMPLE \cite{pathsample} and the disconnectivity trees were drawn using disconnectionDPS \cite{disconnectionDPS}.

\subsection{Financial support and sponsorship}
The work is partially supported by NSF-DMS 1720002.
\url{https://www.nsf.gov/div/index.jsp?div=DMS}



\subsection{Conflicts of interest}

All authors declared that there are no conflicts of interest.

\subsection{Ethical approval and consent to participate}

Not applicable.

\subsection{Consent for publication}

Not applicable.

\subsection{Copyright}

© The Author(s) 2023.

\bibliography{commutetimessvd}

\end{document}